\documentclass[twocolumn]{aastex631}

\newcommand{\ajs}[1]{{{#1}}}
\newcommand{\ajsii}[1]{{{#1}}}

\shorttitle{Lensing Exposure Time Calculator (\textsc{LensingETC})}
\shortauthors{Shajib et al.}
\graphicspath{{./}{figures/}}

\begin{document}

\title{LensingETC: a tool to optimize multi-filter imaging campaigns of galaxy-scale strong lensing systems}

\author[0000-0002-5558-888X]{Anowar J. Shajib}\thanks{NHFP Einstein Fellow}
\affiliation{Department of Astronomy \& Astrophysics, The University of Chicago, Chicago, IL 60637, USA}

\newcommand{\astrod}{{\tt ASTRO\hspace{0.4ex}3D}}

\author[0000-0002-3254-9044]{Karl Glazebrook}
\affiliation{Centre for Astrophysics and Supercomputing, Swinburne University of Technology, Hawthorn, Victoria 3122, Australia}
\affiliation{ARC Centre for Excellence in All-Sky Astrophysics in 3D (\astrod), Australia}

\author[0000-0002-2784-564X]{Tania Barone}
\affiliation{Centre for Astrophysics and Supercomputing, Swinburne University of Technology, Hawthorn, Victoria 3122, Australia}
\affiliation{ARC Centre for Excellence in All-Sky Astrophysics in 3D (\astrod), Australia}

\author[0000-0003-3081-9319]{Geraint F. Lewis}
\affiliation{Sydney Institute for Astronomy, School of Physics, A28, The University of Sydney, NSW 2006, Australia}

\author[0000-0001-5860-3419]{Tucker Jones}
\affiliation{Department of Physics, University of California Davis, 1 Shields Avenue, Davis, CA 95616, USA}

\author[0000-0001-9208-2143]{Kim-Vy H. Tran}
\affiliation{School of Physics, University of New South Wales, Kensington, Australia}
\affiliation{ARC Centre for Excellence in All-Sky Astrophysics in 3D (\astrod), Australia}

\author[0000-0002-3304-0733]{Elizabeth Buckley-Geer}
\affiliation{Fermi National Accelerator Laboratory, P. O. Box 500, Batavia, IL 60510, USA}
\affiliation{Department of Astronomy \& Astrophysics, The University of Chicago, Chicago, IL 60637, USA}
\affiliation{Kavli Institute for Cosmological Physics, University of Chicago, Chicago, IL 60637, USA}

\author[0000-0001-5564-3140]{Thomas E. Collett}
\affiliation{Institute of Cosmology and Gravitation, University of Portsmouth, Portsmouth PO1 3FX, UK}

\author{Joshua Frieman}
\affiliation{Department of Astronomy \& Astrophysics, The University of Chicago, Chicago, IL 60637, USA}
\affiliation{Kavli Institute for Cosmological Physics, University of Chicago, Chicago, IL 60637, USA}
\affiliation{Fermi National Accelerator Laboratory, P. O. Box 500, Batavia, IL 60510, USA}

\author[0000-0003-4239-4055]{Colin Jacobs}
\affiliation{Centre for Astrophysics and Supercomputing, Swinburne University of Technology, Hawthorn, Victoria 3122, Australia}
\affiliation{ARC Centre for Excellence in All-Sky Astrophysics in 3D (\astrod), Australia}



\begin{abstract}

Imaging data is the principal observable required to use galaxy-scale strong lensing in a multitude of applications in extragalactic astrophysics and cosmology. In this paper, we develop Lensing Exposure Time Calculator (\textsc{LensingETC})\footnote{\url{https://github.com/ajshajib/LensingETC}} to optimize the efficiency of telescope time usage when planning multi-filter imaging campaigns for galaxy-scale strong lenses. This tool simulates realistic data tailored to specified instrument characteristics and then automatically models them to assess the power of the data \ajsii{in constraining lens model parameters}. We demonstrate a use case of this tool by optimizing a two-filter observing strategy (in IR and UVIS) within the limited exposure time per system allowed by a \textit{Hubble Space Telescope} (\textit{HST}) Snapshot program. We find that higher resolution is more advantageous to gain constraining power on the lensing observables, when there is a trade-off between signal-to-noise ratio and resolution\ajs{;} e.g., between the UVIS and IR filters of the \textit{HST}. We also find that, whereas a point spread function (PSF) with sub-Nyquist sampling allows the sample mean for a model parameter to be robustly recovered for both galaxy--galaxy and point-source lensing systems, a sub-Nyquist sampled PSF introduces a larger scatter than a Nyquist sampled one in the deviation from the ground truth for point-source lens systems.

\end{abstract}

\keywords{Strong gravitational lensing --- Astronomy software --- Observation methods}


\section{Introduction} \label{sec:intro}

Strong gravitational lensing is the phenomenon where a background source is multiply imaged \ajs{and magnified} through the gravitational deflection of photons by a massive deflector in the foreground. Galaxy-scale strong lenses are useful probes of the mass distribution in galaxies at the intermediate redshift ($0.1 \lesssim z \lesssim 1.5$). Thus, such systems find a multitude of applications in galaxy evolution and cosmology \citep[see, e.g.,][]{Treu10b, Treu16b}.

Imaging data is essential to model the mass distribution in the deflector, either only from the lensed image positions, and/or by reconstructing the flux distribution of the source galaxy. It can potentially be advantageous to use multi-band imaging data in lens modeling instead of single-band data with the same exposure time. If the background galaxies are star-forming galaxies with blue clumps, then the lensed arcs will have more structure in the 
bluer bands, which can provide more constraints for the lens model. However, both the source and the deflector galaxy \ajsii{would have higher signal-to-noise ratio (SNR) per unit angular area given the same exposure time in redder bands for their typical redshifts and spectral energy distributions (specially for space telescopes)}. Furthermore, the point spread function (PSF) differences between the bands can also impact the constraining power of the imaging data. For example, the \textit{Hubble Space Telescope} PSF is not Nyquist sampled \ajs{-- i.e., the full-width-half-maximum of the PSF does not span at least two pixels --}  in the infra-red (IR) channel, however it does in the ultra-violet-visual (UVIS) channel. In addition to lensing features, color information of the deflector is often necessary in applying the strong lensing constraints for specific astrophysical applications. \ajsii{The trade-off between image resolution and the SNR of the UVIS and IR bands is not obvious, especially given the complex structures of features seen in lensing. Thus, to optimize a multi-band observation strategy for efficient usage of limited telescope time, the only way forward is to simulate data to evaluate the achievable uncertainties of key lens model parameters.}

In this paper, we describe a \ajs{new} tool to simulate a sample of galaxy-scale strong lenses corresponding to a provided set of instrument specifications, and then to extract lens model parameter uncertainties from this simulated data (Section \ref{sec:software}). Multiple observing scenarios -- combinations of exposure times and filters -- can be tested
in this tool for comparison and selection of the most efficient strategy. Using this, we test various observing strategies to maximize the constraining power of imaging data obtained from one truncated orbit of a \textit{HST} Snapshot (SNAP) program (Section \ref{sec:snap_test}). Furthermore, we investigate any potential systematic impact of the sampling resolution of the PSF in the lens modeling, both for galaxy--galaxy and point-source lens systems (Section \ref{sec:psf_experiment}). Finally, we discuss our results and summarize the paper in Section \ref{sec:discussion}.

\section{Description of software program} \label{sec:software}

In this section, we describe our software program Lensing Exposure Time Calculator (\textsc{LensingETC}). \textsc{LensingETC} uses \textsc{lenstronomy} \citep{Birrer18, Birrer21b} to simulate and model the mock imaging data. 
\ajsii{However, the user interface (UI) of our program does not require a user to have any experience in using \textsc{lenstronomy} as the UI only requires specifications of the instrument and the observation.\footnote{An easy-to-follow example is provided as a \textsc{jupyter} notebook in the GitHub repository: \url{https://github.com/ajshajib/LensingETC}.} As a result, a user will be able to forecast lens model parameter uncertainties and optimize observing strategies using our software program with minimal setup time ($\sim$30 minutes)  required.}
We describe the simulation of mock lenses in Section \ref{sec:simulation}, and modeling of the mock lenses in Section \ref{sec:modeling}.

\subsection{Simulating mock lens systems} \label{sec:simulation}

\subsubsection{Instrument specifications}

The simulation module requires specifications of the following instrument/image properties for each of the adopted filters: read noise, CCD gain, sky brightness, magnitude zero point, pixel scale, number of pixels in simulated image,
 PSF (can be supersampled or non-supersampled), and cosmic ray event rate (optional). Additionally, a set of observing strategies can be provided to the simulation that specifies the number of exposures per filter and the integration time of a single exposure. The user needs to already account for instrument overhead before providing the exposure times to \textsc{LensingETC}. If a cosmic ray event rate is provided, the different exposures have different simulated realizations of cosmic-ray-hit pixels, and the total exposure time per pixel is computed by disregarding the cosmic-ray-hit pixels in the different exposures. Then, Poisson noise is added to the simulated images based on the effective exposure time per pixel, and background noise and read noise is also applied to the final simulated image.

\subsubsection{Lens galaxy mass and light} \label{sec:lens_profiles}

We use an elliptical power-law mass profile to describe the mass distribution of the lens galaxy \citep{Barkana98, Tessore15}. The convergence \ajs{-- i.e., the surface mass density normalized by the critical density of lensing --} of this mass profile is given by
\begin{equation}
	\kappa_{\rm PL}(x, y) \equiv \frac{3 - \gamma}{2} \left[ \frac{\theta_{\rm E}}{\sqrt{q_{\rm m} x^2 + y^2/q_{\rm m}}} \right]^{\gamma - 1},
\end{equation}
where $\gamma$ is the logarithmic slope, $\theta_{\rm E}$ is the Einstein radius, and $q_{\rm m}$ is the axis ratio. We use the S\'ersic function to describe the light distribution of the lens galaxy, which is given by
\begin{equation}
	I (x, y) \equiv I_{\rm eff} \exp \left[-b_n \left\{\left(\frac{\sqrt{q_{\rm L} x^2 + y^2/q_{\rm L}}}{R_{\rm eff}} \right)^{1/n_{\rm s}} - 1 \right\} \right],
\end{equation}
where $R_{\rm eff}$ is the effective radius along the intermediate axis, $I_{\rm eff}$ is the amplitude at $R_{\rm eff}$, and $n_{\rm s}$ is the S\'ersic index \citep{Sersic68}. \ajs{Additionally,} $b_n$ is a normalizing factor so that $R_{\rm eff}$ becomes the half-light radius. We fix $n_{\rm s} = 4$, which is a typical value for massive elliptical galaxies \citep[e.g.,][]{Tasca11}, as massive ellipticals are the most common type of deflector galaxies in galaxy-scale strong lenses \citep[e.g.,][]{Bolton06, Gavazzi12}. The total apparent magnitude of the lens galaxy can be provided to the simulation module, which is then converted to $I_{\rm eff}$ in the flux unit of electron s$^{-1}$.

\begin{deluxetable}{lll}
\tablecaption{\label{tab:lens_parameters}
Lens galaxy model parameter specifications}
\tablewidth{0pt}
\tablehead{
\colhead{Parameter} & \colhead{Description} & \colhead{Distribution} \\
}
\startdata
$\gamma$ & Power-law exponent & $\mathcal{U}(1.9, 2.1)$ \\
$\theta_{\rm E}$ & Einstein radius & $\mathcal{U}(1\farcs2, 1\farcs6)$ \\
$q_{\rm m}$ & Mass axis ratio &  $\mathcal{U}(0.7, 0.9)$ \\
$\varphi_{\rm m}$ & Mass position angle & $\mathcal{U}(-90\deg, 90\deg)$ \\
$\gamma_1$ & External shear & $\mathcal{U}(-0.08, 0.08)$ \\
$\gamma_2$ & External shear & $\mathcal{U}(-0.08, 0.08)$ \\
$q_{\rm L}$ & Light axis ratio & Joint with $q_{\rm m}$ \\
$\varphi_{\rm L}$ & Light position angle & Joint with $\varphi_{\rm m}$ \\
\enddata
\tablecomments{$\gamma_1$ and $\gamma_2$ are external shear parameters in the Cartesian coordinate system that are related to the external shear magnitude $\gamma_{\rm ext}$ and $\varphi_{\rm ext}$ as $\gamma_{1} \equiv \gamma_{\rm ext} \cos (2\varphi)$ and $\gamma_2 \equiv \gamma_{\rm ext} \sin (2\varphi)$.}
\end{deluxetable}

\subsubsection{Source galaxy light} \label{sec:source_profiles}

\begin{figure}
	\includegraphics[width=\columnwidth]{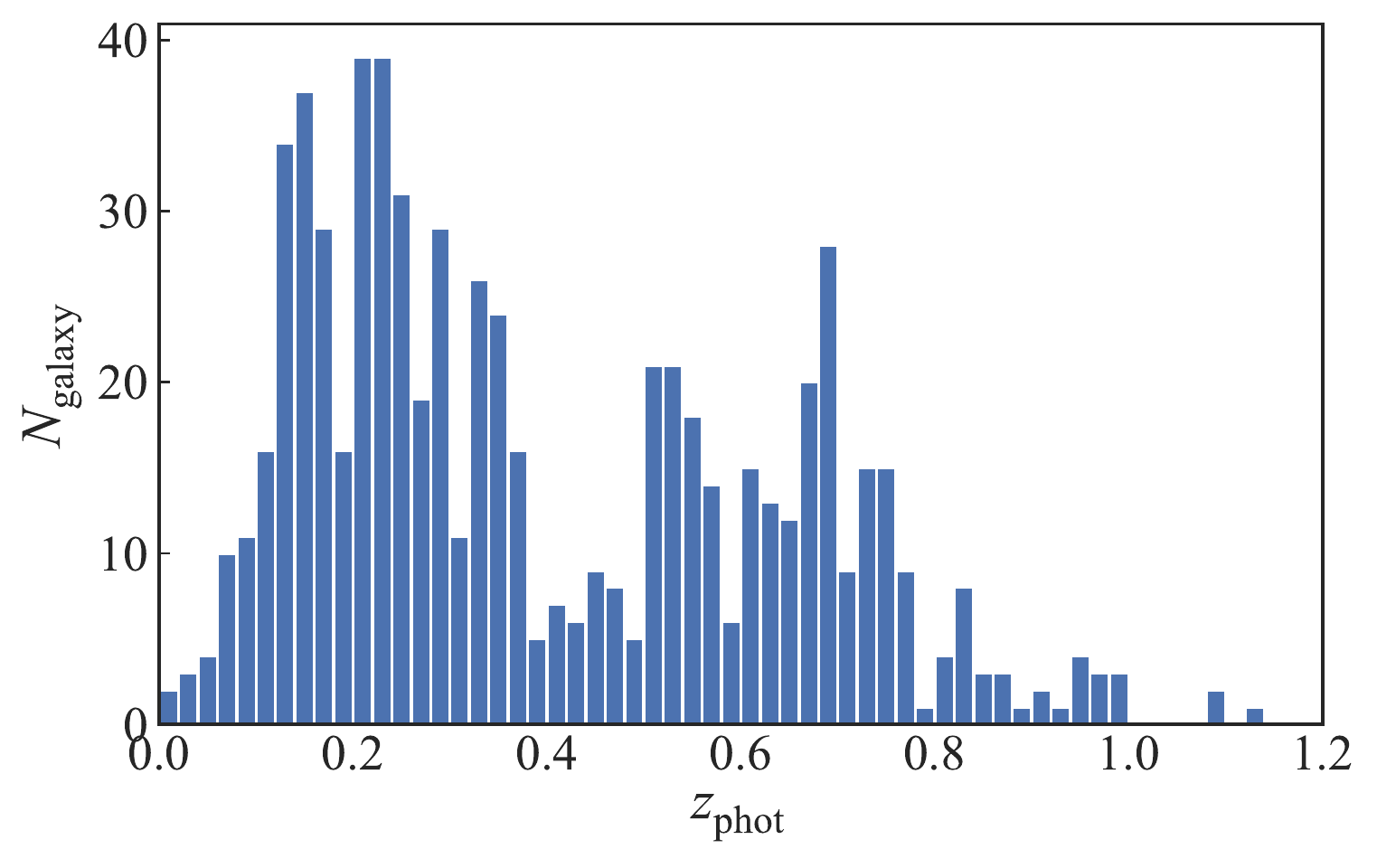}
	\caption{\label{fig:redshift_distribution}
	Distribution of photometric redshifts of the 699 spiral and disky galaxies selected from the Galaxy Zoo catalogue.
	}
\end{figure}

We use light profiles of real spiral and disky galaxies from the \textit{HST} legacy imaging as the source galaxy light distribution. We select 1,117 spiral and disky galaxies from the Galaxy Zoo catalog of 120,000 galaxies that are morphologically classified through crowd-sourcing \citep{Willett17}. \ajsii{To select spiral and disky galaxies}, we use these criteria: \texttt{t01\_smooth\_or\_features\_a01\_smooth\_flag = True} or \texttt{t01\_smooth\_or\_features\_a02\_features\_or\_disk\_flag = True}, \texttt{t06\_odd\_a02\_no\_weighted\_fraction > 0.85} \citep[for definitions of the criteria, see][]{Willett17}. The selected images are from either of the surveys: the All Wavelength Extended Groth Strip International Survey \citep[AEGIS;][]{Georgakakis07}, Galaxy Evolution from Morphologies and SEDs \citep[GEMS;][]{Caldwell08}, and the Great Observatories Origins Deep Survey \citep[GOODS;][]{Giavalisco04}. The photometric redshifts of the selected galaxies range from 0.01 to \ajsii{1.33 (see Figure \ref{fig:redshift_distribution}). Although typical source galaxies of strong lensing systems have a mean redshift of $\sim$2, we only use the morphologies of the selected galaxies in the simulation and not their colors, which needs to be set by the user appropriately. Thus, the redshift distribution of these galaxies is unimportant for the purpose of our software program.}

\begin{figure*}
	\begin{center}
	\includegraphics[width=1\textwidth]{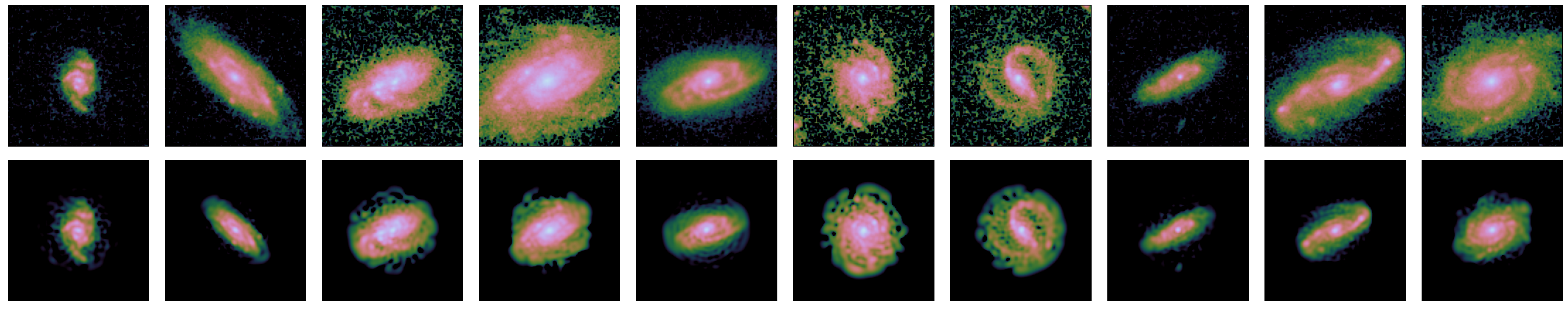}
	\caption{ \label{fig:source_galaxies} 
	\textbf{Top row}: \textit{HST} legacy imaging of spiral galaxies from the Galaxy Zoo catalog. \textbf{Bottom row}: Reconstruction of the light distribution based on a basis set of shapelets (i.e., 2D Gauss--Hermite polynomial) with maximum polynomial order $n_{\rm max} = 50$. The shapelet-based reconstructions are rescaled to the same characteristic size. 
	}
	\end{center}
\end{figure*}

Using the RA and \ajs{Dec} of the selected galaxies, we retrieve $5\arcsec \times 5\arcsec$ cutouts from the \textit{HST} legacy archive. These images are from the ACS camera in the F606W filter. Based on visual inspection of the downloaded images, we further remove images with noticeable cosmic ray hits, systems near the edge of the cutout, systems that are too faint or too small (i.e., only covering $\lesssim10$ pixels), systems with very irregular shape or light distribution not resembling a spiral or disky galaxy, and systems that are too large extending beyond the cutout size. After this quality control procedure, we end up with 699 images cutouts of disky and spiral galaxies (for a few examples, see Figure \ref{fig:source_galaxies}).

We \ajsii{extract the morphological information of the source galaxies by fitting} a basis set of shapelets -- i.e., 2D Gauss--Hermite polynomials -- to these galaxy light profiles \citep{Refregier03}.\footnote{\ajs{Although \textsc{lenstronomy} allows using a pixelated flux profile as the source galaxy light distribution through the \texttt{`INTERPOL'} profile \citep{Wagner-Carena22}, we find the shapelet-based description to be more advantageous due to its flexibility in rescaling the galaxy size and in tuning the degree of clumpy-ness in the simulation.}} \ajsii{Extracting the morphological information in the parametric form of shapelets allows us to easily scale the sizes of these galaxies in our simulation.} We choose the highest polynomial order $n_{\rm max} = 50$. Thus, the total number of shapelet components (i.e., the number of free linear parameters) in the basis set is $(n_{\rm max} + 1)(n_{\rm max} + 2)/2 = 1326$. We select the most appropriate shapelet scale parameter ($\beta$) for this fitting by finding the minimum $\chi^2$ fit out of multiple fits with gradually increasing $\beta$ values from 0\farcs025 to 1\arcsec with a uniform step size of 0\farcs025. Once the best fit amplitudes of the $1326$ shapelet components are obtained with an optimal $\beta$ parameter, then the $\beta$ parameter can be randomly set in the simulation to enlarge or shrink the size of the galaxy while retaining the morphological structure of the galaxy, which is set by the relative amplitudes of the shapelet components. Figure \ref{fig:source_galaxies} illustrates some examples of the real \textit{HST} images of the source galaxies and their shapelet-based reconstructions.
For ease of use, the user can select particular galaxies to use as the source from our list of 699 galaxies instead of relying on random selection, or can directly provide shapelet coefficients if other galaxy morphologies not included in our list are desired.

Additionally, a point source can be added at the center of the source galaxy to simulate the presence of a quasar or a supernova. 
\ajs{Although a supernova may not necessarily be at the center of the host galaxy, the exact position of the supernova within the galaxy does not impact the statistical constraining power of the imaging data at the sample level given the random positioning of the source galaxy with respect to the caustic curve.}
The user is required to provide distributions of the total magnitudes corresponding to lens galaxy, source galaxy, and the point source for each adopted filter.\footnote{\ajsii{For some example values to use as the magnitude distributions  that are consistent with those of the lens sample presented by \citet{Shajib19}, see Sections \ref{sec:snap_test} and \ref{sec:psf_experiment}.}}

\subsection{Modeling of the simulated lenses} \label{sec:modeling}

\ajsii{Our software program automatically sets up the lens model specifications to optimize the model with \textsc{lenstronomy}. The modeling of the simulated images is performed in the same way as done for real imaging data \citep[e.g.,][]{Shajib19, Shajib21}. The mass and light profiles that are optimized in the lens modeling have the same parameterization as in Section \ref{sec:lens_profiles} and \ref{sec:source_profiles}.} The model parameter uncertainties are obtained from Markov Chain Monte Carlo (MCMC) sampling using the software \textsc{emcee} \citep{Goodman10, Foreman-Mackey13}. In the modeling, the user can choose to provide a different or lower resolution PSF than the one used for simulation for any of the specified filters to mimic a real world modeling scenario. The user can also choose different $n_{\rm max}$ parameters in simulation and modeling, since usually $n_{\rm max} \leq 10$ is sufficient to model lensing systems for the resolution of current high-end facilities such as the \textit{HST} \citep[e.g.,][]{Shajib19}.

\section{Optimal observing strategy for \textsc{HST} SNAP program} \label{sec:snap_test}

\ajs{Here}, we perform a comparison test between different observing strategies of galaxy--galaxy lens systems for the \textit{HST} SNAP program \ajsii{GO-16773  that aims to image strong lensing systems discovered in Dark Energy Survey data \citep{Jacobs19, Jacobs19b}. The observations in a SNAP program are limited to $\lesssim$45 minutes including overheads, as targets from these programs are used to fill in scheduling gaps in between targets from General Observer (GO) programs.} 

\subsection{Test setup}

\begin{figure}
	\includegraphics[width=\columnwidth]{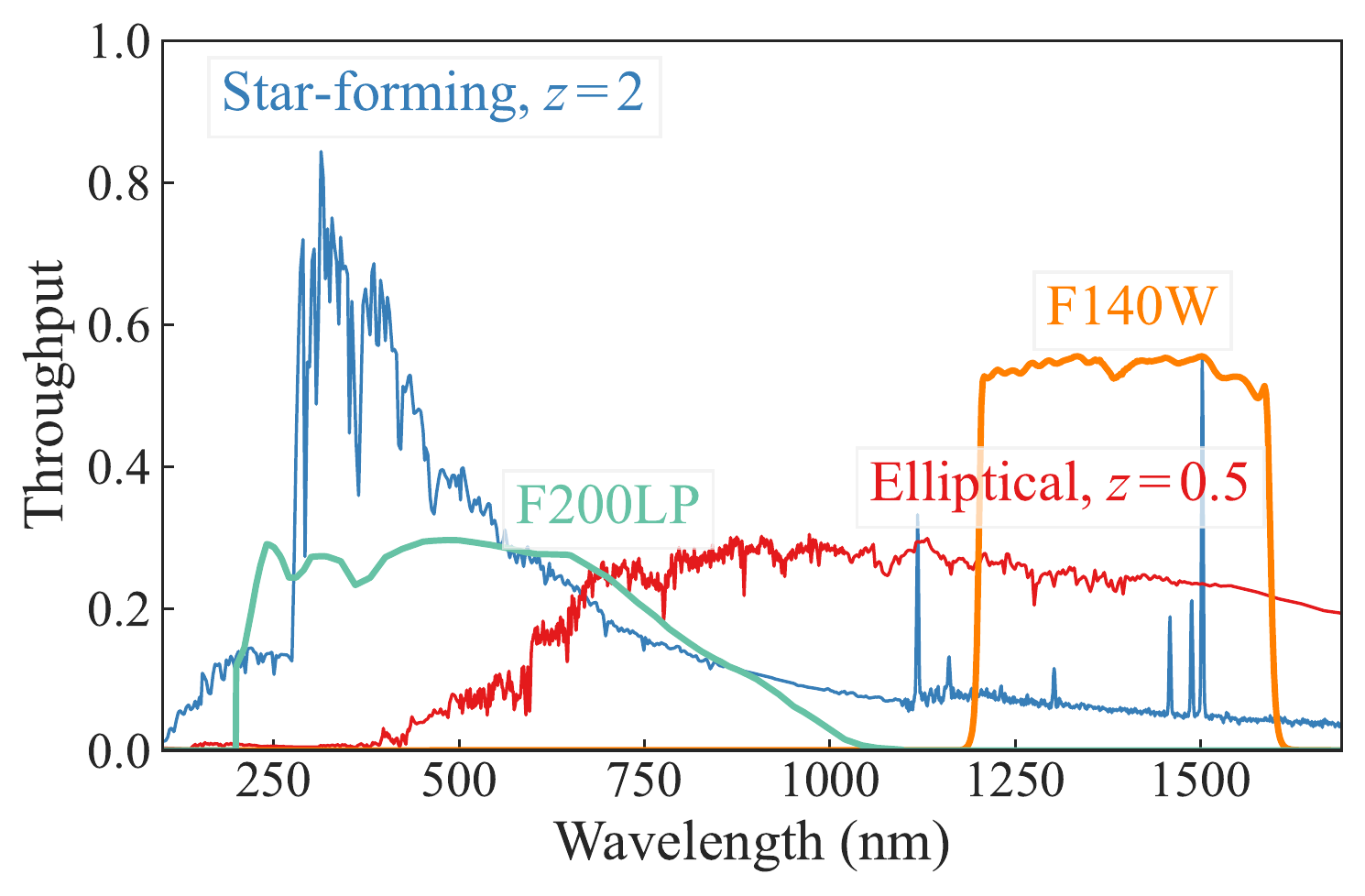}
	\caption{\label{fig:filter_throughput}
	\ajs{Throughput of the F200LP (emerald line) and F140W (orange line) filters of the \textit{HST} WFC3. The spectral energy distributions of a star-forming galaxy (blue line) at $z=2$ and an elliptical galaxy (red line) at $z=0.5$ are also illustrated with arbitrary amplitude scaling. Each of the two adopted filters maximizes the relative contribution from either of the deflector or the source galaxy to the total light.}
	}
\end{figure}

We choose two filters in our setup: F140W (IR) and F200LP (UVIS). Whereas deflector galaxies -- which are typically red ellipticals -- are brighter in the IR band, the source galaxies -- which are typically star forming galaxies -- have more structure from the star-forming regions in the bluer UVIS band (see Figure \ref{fig:filter_throughput}). \ajsii{We choose these two filters, because they are wide filters that allow maximize the collected photons and thus the SNR given the exposure time constrain of a SNAP program. Although the PSF in wide filters has relatively stronger color dependence than the narrow ones, we find that this color dependence is insignificant in the context of lens modeling \citep[see also,][]{Shajib21}.}


We choose six  cases of exposure time allocation that range from allocating the maximum time for the F140W filter to allocating the maximum time for the F200LP filter (see Table \ref{tab:snap_experiment_cases}). We have already accounted for the instrument overheads using the \textit{HST} Astronomer's Proposal Tool (APT) and chosen the exposure sequences so as to minimize the overhead within the observing time limit.

\begin{deluxetable*}{cccc}
\tablecaption{\label{tab:snap_experiment_cases}
Different observing strategies for a \textit{HST} SNAP program in our test setup.}
\tablewidth{0pt}
\tablehead{
\colhead{Case} & \colhead{F140W exposure time} & \colhead{F200LP exposure time} & \colhead{Total SNAP visit time including overheads}\\
& (s) & (s) & (s)
}
\startdata
1 & $4\times350$ & -- &  1905 \\
2 & $3\times300$ & $1\times300$ & 1773 \\
3 & $2\times300$ & $2\times300$ & 1881 \\
4 & $3\times200$ & $2\times300$ & 1901 \\
5 & $1\times300$ & $2\times450$ & 1861 \\
6 &  -- & $2\times650$ & 1819 \\
\enddata
\end{deluxetable*}

We choose the pixel sizes 0\farcs08 for the F140W filter and 0\farcs04 for the F200LP  filter to mimic the \ajsii{pixel scales} of drizzled images \citep{Shajib19}. For the F140W filter, we use a supersampled PSF with resolution factor of 3 in the simulation, but for modeling we use a non-supersampled PSF. For the F200LP filter, the same non-supersampled PSF is used for both simulation and modeling. However, the UVIS pixel size of 0\farcs04 in the PSF already Nyquist samples the \textit{HST} PSF. For both simulation and modeling, we 
choose $n_{\rm max} = 4$ and $n_{\rm max} = 8$ for the F140W and F200LP filters, respectively. We choose a lower $n_{\rm max}$ for the F140W filter, since the resolution is lower for this filter and thus the finer structures in the lensed arcs would be smeared away making higher $n_{\rm max}$ unnecessary. In the simulation, we use the following magnitude and color distributions that are consistent with the lens systems presented in \citet{Shajib19}:
\begin{itemize}
	\item Lens galaxy: $m_{\rm F140W} \sim \mathcal{N}(18.5, 0.5)$, $m_{\rm F200LP} - m_{\rm F140W} \sim \mathcal{N}(2.7, 0.2)$,
	\item Unlensed source galaxy: $m_{\rm F140W} \sim \mathcal{N}(20.80, 0.08)$, $m_{\rm F200LP} - m_{\rm F140W} \sim \mathcal{N}(2.0, 0.2)$.
\end{itemize}
We add the the effect of cosmic rays in the simulated exposures. We set the cosmic ray event rate to be $1.2\times10^{-4}$ s$^{-1}$ arcsec$^{-2}$ in the F140W filter and $2.4\times10^{-4}$ s$^{-1}$ arcsec$^{-2}$ in the F200LP filter. These values are estimated by counting cosmic ray hits within a $10\arcsec\times10\arcsec$ area in real \textit{HST} exposures. The event rate in the F140W filter is smaller than the F200LP filter, because not every cosmic ray hit is catastrophic in the IR detector as \ajsii{non-destructive readout is possible in this detector}. However, the event rate in the F140W filter accounts for the presence of random dead or hot pixels in the exposures. 

\begin{figure*}
\centering
	\includegraphics[width=0.9\textwidth]{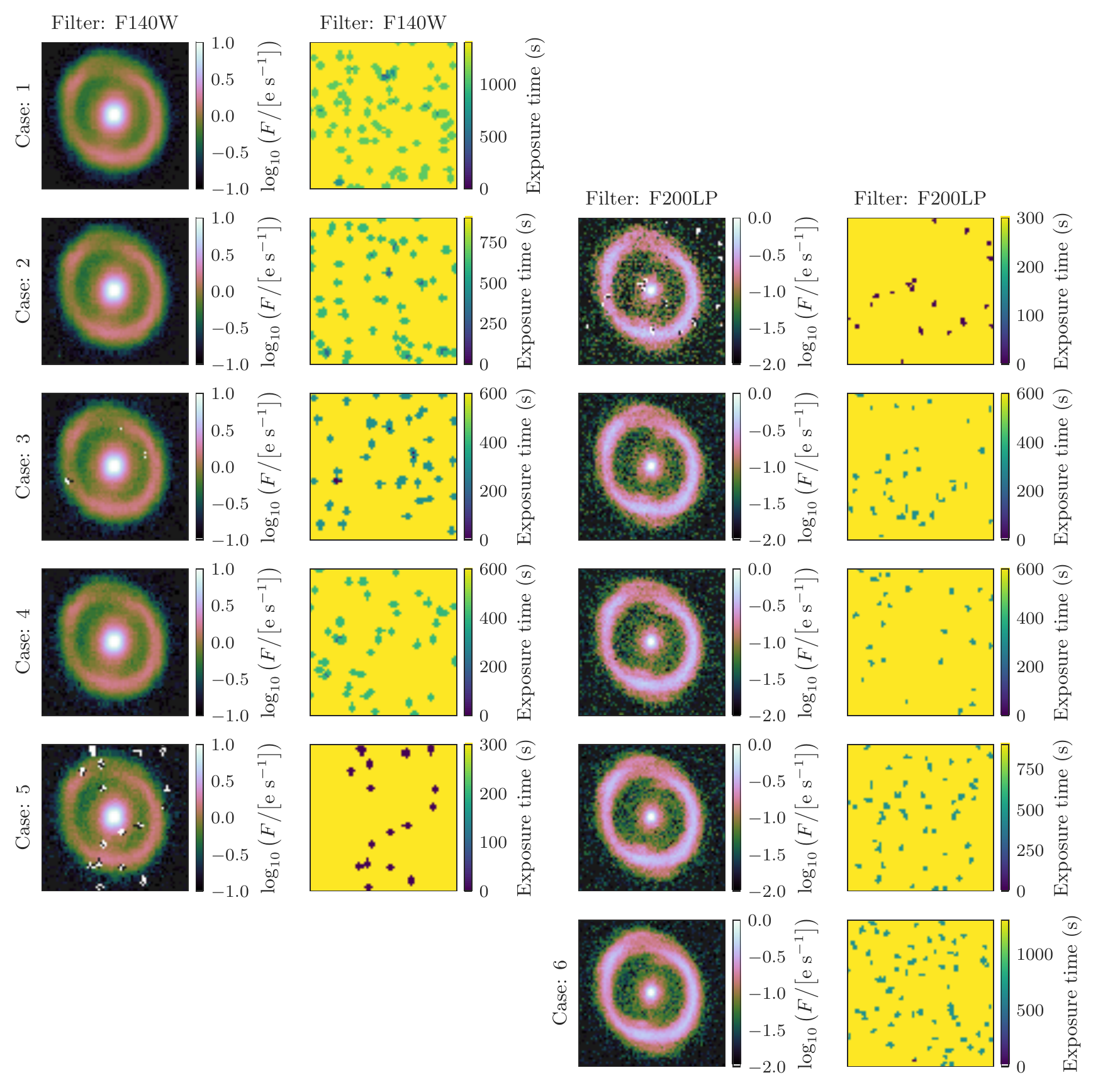}
	\caption{ \label{fig:simulated_systems_snap}
	Example of a simulated lens system in the six observing scenarios and two filters for a \textit{HST} SNAP program. The first and third columns illustrate the simulated imaging data, and the second and fourth columns illustrate the exposure time map after accounting for the impact of cosmic ray hits. A montage of the full sample of 20 lenses is given in Figure \ref{fig:simulated_lens_montage}.
	}
\end{figure*}

We simulate 20 lensing systems for each of the six observing scenarios and obtain the model parameter uncertainties for each case from MCMC chains. Figure \ref{fig:simulated_systems_snap} illustrates \ajs{several} examples of the simulated images and the corresponding exposure maps with random realizations of cosmic ray hits.

\subsection{Result}

\begin{figure*}
	\includegraphics[width=0.5\textwidth]{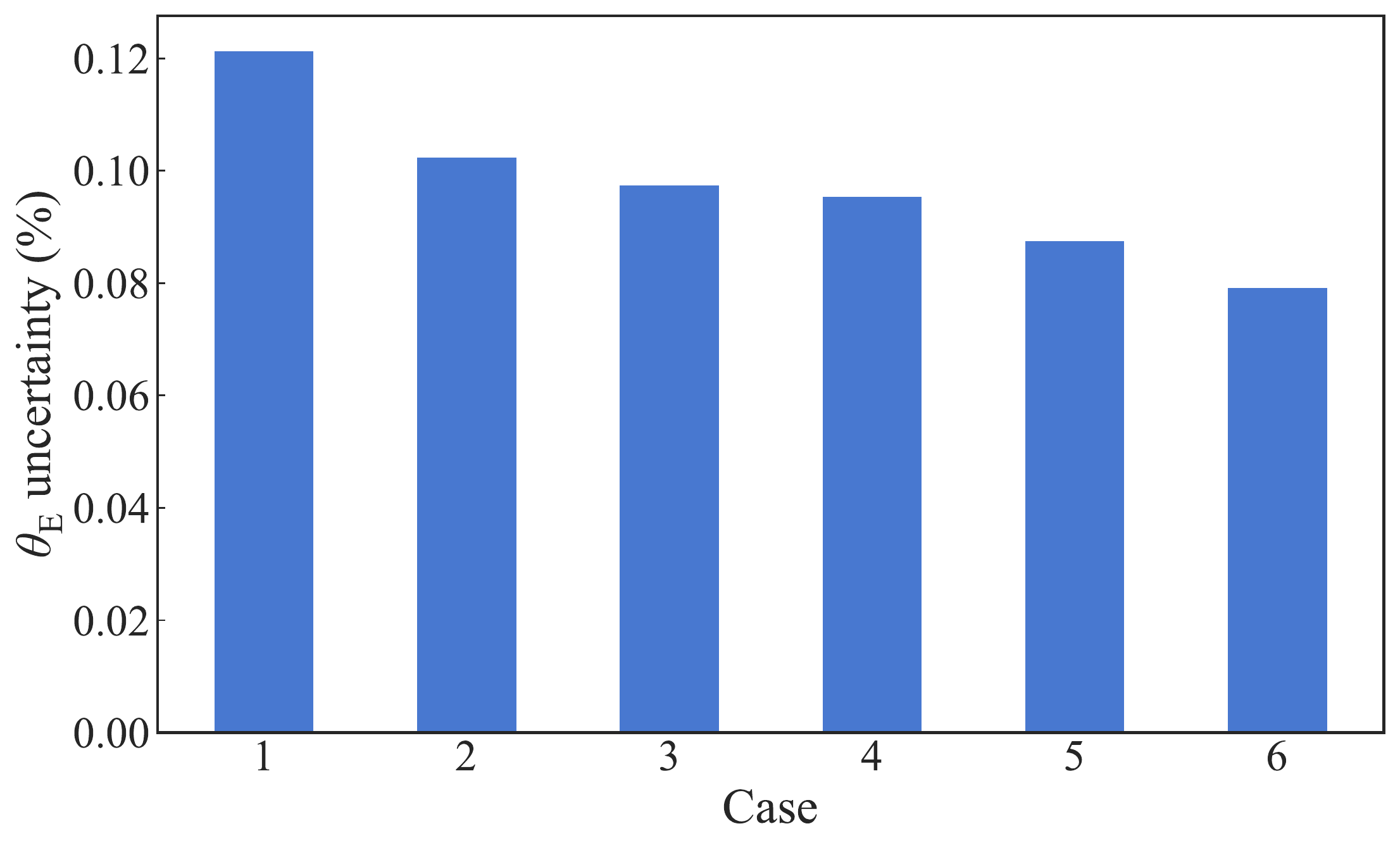}
	\includegraphics[width=0.5\textwidth]{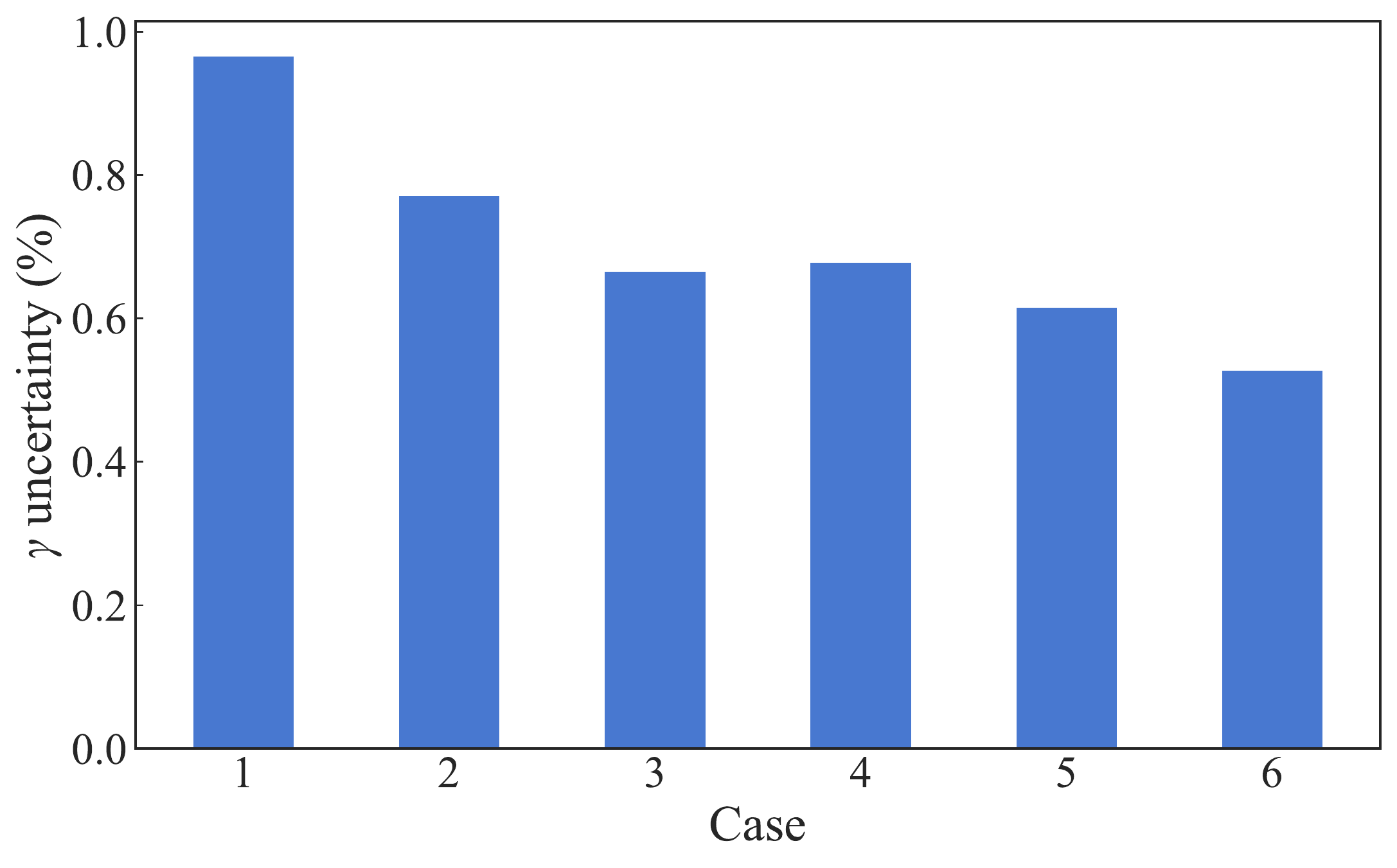}
	\caption{\label{fig:hst_snap_param_comparison}
	Comparison of parameter uncertainties between the six observing strategies for \textit{HST} SNAP program: for Einstein radius $\theta_{\rm E}$ in the left-hand panel, and for the logarithmic slope parameter $\gamma$ in the right-hand panel. The six strategies distribute $\sim$2000 s of total observing time of one \textit{HST} truncated orbit between two filters: from allocating all the time for the F140W filter in Case 1 to allocating all the time for the F200LP filter in Case 6. Since the F200LP band allows higher resolution than the F140W band, higher image resolution is more constraining for lens models than higher SNR for typical strong lensing systems.
	}
\end{figure*}

In Figure \ref{fig:hst_snap_param_comparison}, we compare the posterior uncertainties of the Einstein radius $\theta_{\rm E}$ and the power-law slope $\gamma$ between the observing strategies. We choose these two parameters for comparing the constraining power of the data for different strategies, since they are central lens model parameters pertaining to the deflector galaxy.\footnote{Although the primary direct observables in the imaging data are the Einstein radius $\theta_{\rm E}$ and the quantity $\theta_{\rm E} \alpha^{\prime\prime} (\theta_{\rm E}) / \left[1 - \kappa(\theta_{\rm E})\right]$ with $\alpha^{\prime \prime}$ being the double derivative of the deflection angle \citep{Kochanek20, Birrer21}, for the power-law model the latter quantity simply becomes $\gamma - 2$. Since we use a power-law model as the ground truth to simulate the data, adopting $\gamma$ as the observable quantity serves the purpose of assessing the constraining power of the data \ajsii{on the lensing mass distribution}.} We find that the constraining power of the data increases for both $\theta_{\rm E}$ and $\gamma$ as \ajsii{a greater fraction of a fixed time} is allocated for the higher resolution UVIS filter. In other words, for realistic magnitudes and colors of typical deflector and source galaxies, higher image resolution is more advantageous than higher SNR to constrain the lens models.

Based on this result, we adopt Case 4 as the observing strategy for our \textit{HST} SNAP program GO-16773. \ajsii{The science cases of this program require measured colors of the deflector and the source galaxies, thus Case 1 and 6 are not considered as both of these use only a single filter. We only included these cases in our experiment to illustrate the edge cases.} Although both Case 3 and 4 allocate the same total exposure time for the F140W filter, Case 4 allows for a more artifact-free drizzled image by dividing the total exposure time into 3 exposures instead of 2 (see Figure \ref{fig:simulated_systems_snap}). \ajsii{Although Case 5 provides better constraints than Case 4, Case 5 is not chosen for the same reason above, as the cosmic-ray-hit and dead pixels lead to considerably degraded IR photometry, which is not desirable to meet the requirements of the science cases.}

\section{Impact of PSF sampling in modeling} \label{sec:psf_experiment}

In this section, we perform a test using \textsc{LensingETC} on the impact of the PSF   sampling resolution in lens modeling. A supersampled PSF, or even a Nyquist-sampled PSF, is not always available \textit{a priori} for lens modeling. For systems with point sources, the PSF is usually reconstructed from an initial estimate based upon nearby stars \citep{Chen16, Birrer19b}. For galaxy--galaxy strong lenses, an estimate of the PSF -- either from nearby stars or from simulation (e.g., \textsc{tinytim} for \textit{HST}) -- is used for modeling. The PSF is not Nyquist sampled in the IR channel of \textit{HST}, for example. Here, we investigate whether such sub-optimally sampled PSFs systematically impact the lens model posterior. As the PSF is integral in lens modeling to extract the lensing information, this investigation will motivate if additional procedures are necessary to obtain supersampled and accurate PSFs as part of an imaging program.

\subsection{Test setup}
We choose 4 scenarios that are combinations of galaxy--galaxy and point-source lens systems, and supersampled and non-supersampled PSFs (see Table \ref{tab:psf_experiment_cases}).


\begin{deluxetable}{ccc}
\tablecaption{\label{tab:psf_experiment_cases}
Test scenarios with different lens system types and PSF supersampling.}
\tablewidth{0pt}
\tablehead{
\colhead{Scenario} & \colhead{System type} & \colhead{PSF}
}
\startdata
1 & galaxy--galaxy lens system & supersampled \\
2 & galaxy--galaxy lens system & non-supersampled \\
3 & point-source lens system & supersampled \\
4 & point-source lens system & non-supersampled \\
\enddata
\end{deluxetable}

For all these scenarios, we only simulate single-filter images with the instrument specifications corresponding to \textit{HST} Wide-Field Camera 3 (WFC3) F140W filter. A supersampled PSF with resolution factor of 3 is used to simulated the images in all the scenarios. We adopt the total exposure time of 8$\times$275 s per system \citep[achievable from one orbit of a \textit{HST} GO program, e.g.,][]{Shajib19}. For the scenarios with non-supersampled PSFs for modeling, we degrade the supersampled PSF through interpolation to match the PSF pixel size with the imaging pixel size. We adopt a 50\% flux uncertainty for the PSFs used in modeling \ajs{the point-source lens systems}, which is a realistic uncertainty level as found in the experiment done by \citet{Ding21}. The simulated point-source lens systems have the same background galaxy as in the simulated galaxy--galaxy lens systems, and we add a point source at the center of the background galaxy with an unlensed magnitude sampled from $m_{\rm F140W} \sim \mathcal{U}(20, 21)$. The deflector and host galaxy magnitude distributions are the same as in Section \ref{sec:snap_test}. We simulate 20 lens systems for each scenario. \ajs{For the sample of point-source lenses, we have 5 quadruple-image systems and 15 double-image systems out of the 20 from random positioning of the source centroid on the source plane.}

\subsection{Result}

\begin{figure}
	\includegraphics[width=0.5\textwidth]{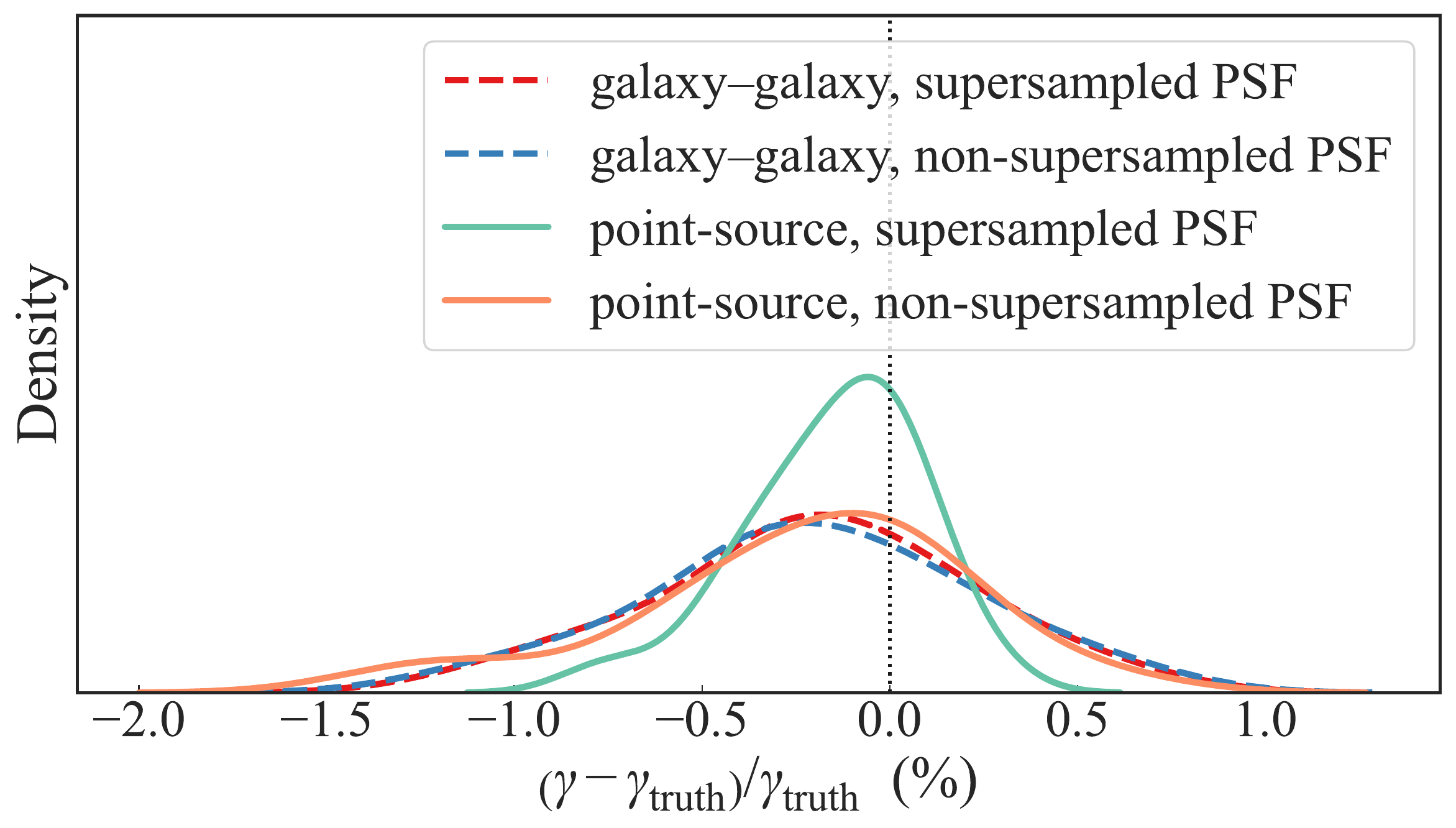}
	\caption{\label{fig:psf_experiment}
	Deviations in power-law slope parameter $\gamma$ for galaxy--galaxy and point-source lenses modeled with and without supersampled PSF. The illustrated distributions correspond to kernel density estimates from 20 lens systems for each case. For galaxy--galaxy lens systems (dashed lines), the PSF supersampling does not noticeably impact the distribution of deviations. However, for point-source lenses, modeling with the non-supersampled PSF creates a larger scatter than that with the supersampled PSF. In all cases, the mean deviation at the sample level is within 0.25\% from zero.
	}
\end{figure}

Figure \ref{fig:psf_experiment} compares the systematic deviations in the modeled $\gamma$ parameter in the 4 test scenarios. For all the scenarios, the mean deviation at the sample level is within 0.25\% from zero. For galaxy--galaxy lens systems, there is no significant difference in the distributions of the deviations between the scenarios having supersampled and non-supersampled PSFs. However, for systems with point sources, the case with the non-supersampled PSF has a larger scatter in the $\gamma$ parameter than the case with the supersampled PSF. This is due to \ajs{the fact} that the accuracy of the point-source positioning depends on the PSF resolution and the positions of the modeled point-sources impact the lens model posterior.

\section{Discussion and Summary} \label{sec:discussion}

In this paper, we present a simulation tool, \textsc{LensingETC}, to optimize observation strategies of galaxy-scale strong lensing systems to achieve maximum constraining power from the imaging data. This tool is built on the lens modeling software program \textsc{lenstronomy}. We provide \ajsii{an easily accessible UI} that does not require any prior experience in using \textsc{lenstronomy}. \ajsii{The user only needs to provide instrument and observation specifications, and apparent magnitude distributions of the deflector and the lensed objects. Our tool then simulates mock imaging data and perform lens modeling with the mock data to obtain model parameter uncertainties.} Thus, our tool allows an investigator to perform the optimization with minimal setup time ($\lesssim$30 minutes) required. This tool can be used to experiment with both galaxy--galaxy and point-source lens systems. For simulated imaging data with 2 filters, running the MCMC chain to obtain the model posterior requires $\sim$2 CPU hours for each combination of lens and observing case\footnote{\ajs{We performed the MCMC sampling on a system with AMD Ryzen 7 3700X 8-core processor with a clock-speed of 3.6 GHz.}}. Thus, for typical imaging programs with $\sim$5 observing cases to optimize from -- such as the one performed in Section \ref{sec:snap_test} -- the optimization procedure takes order of days to run all the MCMC chains for the combinations of simulated lenses and cases.

Although our tool is built for galaxy-scale strong lensing systems, the results can be qualitatively generalized to group-scale and cluster-scale strong lensing systems. The lensing observables captured in the lensed arcs within the imaging data can be generalized for all cases of strong lensing and even up to the regime of weak lensing \citep{Birrer21}. Our tool uses the same lens model to simulate and fit the imaging data, the model posterior only captures the statistical uncertainty without any additional modeling systematics. Thus, the statistical constraining power of the local lensing observables contained in the lensed arcs can be generalized to group- and cluster-scales for a similar range of lensing magnification and source magnitude. However, specific systematics that may potentially arise from particular modeling methods used for group-scale and cluster-scale lenses are not captured in the results of our tool, thus these systematics may be required to be considered before making quantitative forecasts using the results from the tool presented in this paper.

We use our tool to optimize the observation strategy for a \textit{HST} SNAP program with $\lesssim$2000 s as the total observing time per lens system including overhead. We find that higher imaging resolution of the \textit{HST} UVIS channel provides more constraining power than the higher SNR achievable with the IR channel. Thus, the resolved structure in the lensed arcs deliver more lensing information to constrain the lens model than a higher SNR in the flux distribution of the lensed arcs.

We furthermore test the importance of PSF sampling resolution in robust estimation of the lens model parameters. For both galaxy--galaxy and point-source lens systems, the mean deviation of the power-law slope $\gamma$ at the sample level is within 0.25\% from zero for both optimal and non-optimal (i.e., sub-Nyquist) PSF sampling. This deviation is negligible given the typical uncertainty achieved with imaging data from current instruments is $\sim$2--5\% \citep[][Schmidt et al. 2022, in preparation]{Shajib19, Shajib21}. We find that the PSF supersampling resolution has no impact in the modeling galaxy--galaxy strong lensing systems. This is consistent with \citet{Shajib21}, who find no significant difference from using multiple simulated and empirical PSFs with varying spectral energy distribution and resolution in modeling \textit{HST} images of galaxy--galaxy lenses. 

However, we find that a sub-Nyquist sampled PSF creates larger scatter in the deviation of the recovered $\gamma$ parameter from the ground truth than a higher resolution PSF above the Nyquist limit. Such a large scatter is consistent with \citet{Shajib22}, who find that the PSF resolution can significantly ($\sim4\sigma$) shift the best fit $\gamma$ parameter, and a supersampled PSF leads to consistent lens model posteriors even if different modeling softwares are used. Therefore, supersampled PSFs are recommended in lens modeling applications such as the time-delay cosmography that is strongly sensitive to the $\gamma$ parameter and typically uses one lens system at a time. Additional observing procedures, e.g., observing nearby stars or choosing systems with a larger number of nearby stars to fit within the same \textit{HST} frame, may be necessary to create a robust initial PSF estimate with supersampled resolution. However, once a large sample of lens systems are considered the scatter in the $\gamma$ parameter will be averaged out to bring the sample mean closer to the ground truth, i.e., the systematics from sub-Nyquist sampled PSFs can be mitigated at the sample level.

\begin{acknowledgments}
\ajs{We thank Simon Birrer for helpful comments that improved this manuscript.} Support for this work was provided by NASA through the NASA Hubble Fellowship grant
HST-HF2-51492 awarded to AJS by the Space Telescope Science Institute, which is operated by the Association of Universities for Research in Astronomy, Inc., for NASA, under contract NAS5-26555.
TJ gratefully acknowledges funding support from the Gordon and Betty Moore Foundation through Grant GBMF8549, from NASA through grant HST-GO-16773.002-A, and from a Dean’s Faculty Fellowship.
TEC is funded by a Royal Society University Research Fellowship and the European Research Council (ERC) under the European Union’s Horizon 2020 research and innovation programme (LensEra: grant agreement No 945536).
\end{acknowledgments}

%

\vspace{5mm}
\facilities{\textit{HST}, MAST}


\software{
\textsc{lenstronomy} \citep{Birrer18, Birrer21d}, 
\textsc{emcee} \citep{Foreman-Mackey13},
\textsc{numpy} \citep{Oliphant15},
\textsc{scipy} \citep{Jones01},
\textsc{matplotlib} \citep{Hunter07},
\textsc{seaborn} \citep{Waskom14}.
}



\appendix

\section{Simulated lenses}

\begin{figure*}
	\includegraphics[width=\textwidth]{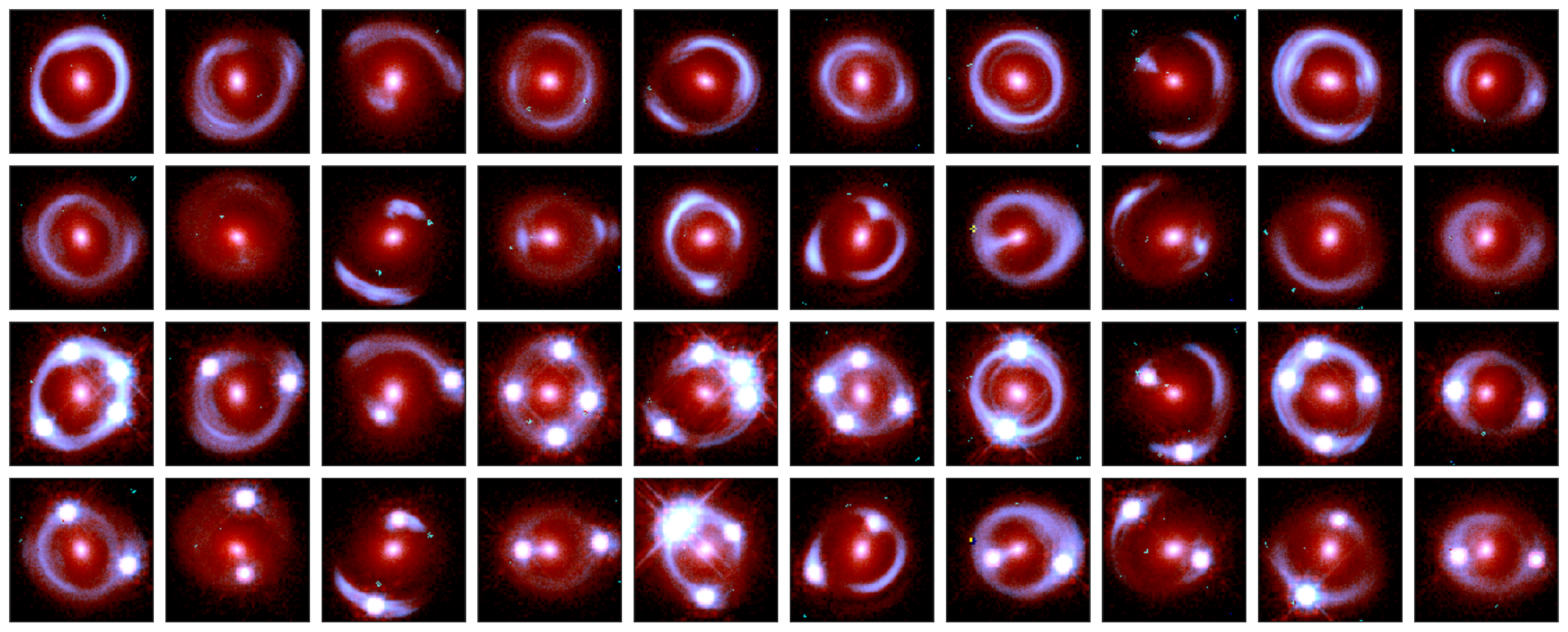}
	\caption{\label{fig:simulated_lens_montage}
	False-color composite images of the simulated galaxy--galaxy (tow 2 rows) and point-source (bottom 2 rows) lens systems from Sections \ref{sec:snap_test} and \ref{sec:psf_experiment}. The RGB images are created from a weighted combination of monochrome images from the F140W and F200LP filters.
	}
\end{figure*}

In Figure \ref{fig:simulated_lens_montage}, we illustrate false-color composites of all the simulated galaxy--galaxy and point-source lenses simulated in Sections \ref{sec:snap_test} and \ref{sec:psf_experiment}.

\bibliography{ajshajib}{}
\bibliographystyle{aasjournal}



\end{document}